\documentclass[11pt,a4paper]{article}

\usepackage{epsfig,amsmath,amssymb,cite}
\def\lp{\left(}
\def\rp{\right)}
\def\lb{\left[}
\def\rb{\right]}
\def\be{\begin{equation}}
\def\ee{\end{equation}}
\begin{document}

\title{Some general aspects of thin-shell wormholes \\
 with cylindrical symmetry} 
\author{Ernesto F. Eiroa$^{1,2,}$\thanks{e-mail: eiroa@iafe.uba.ar}, 
Claudio Simeone$^{2,}$\thanks{e-mail: csimeone@df.uba.ar}\\
{\small $^1$ Instituto de Astronom\'{\i}a y F\'{\i}sica del Espacio, C.C. 67, 
Suc. 28, 1428, Buenos Aires, Argentina}\\
{\small $^2$ Departamento de F\'{\i}sica, Facultad de Ciencias Exactas y 
Naturales,} \\ 
{\small Universidad de Buenos Aires, Ciudad Universitaria Pab. I, 1428, 
Buenos Aires, Argentina}} 
\maketitle

\begin{abstract}

In this article we study a general class of non-rotating thin-shell wormholes with cylindrical symmetry. We consider two physically sound definitions of the flare-out condition and we show that the less restrictive one allows for the construction of wormholes with positive energy density at the throat. We also analyze the mechanical stability of these objects under perturbations preserving the symmetry, proving that previous results are particular cases of a general property. We present examples of wormholes corresponding to Einstein--Maxwell spacetimes.\\

\noindent 
PACS number(s): 04.20.Gz, 04.20.Jb, 04.40.Nr\\
Keywords: Lorentzian wormholes; exotic matter; Einstein--Maxwell spacetimes

\end{abstract}

\section{Introduction}

Wormholes are solutions of the equations of gravitation which imply a nontrivial topology of spacetime, connecting two regions of the universe --or of two different universes-- by a traversable throat. For compact configurations, the throat is a minimal area surface \cite{hovis}; this feature is determined by the fact that the geodesics open up at the throat (for an alternative definition in terms of trapped surfaces see Ref. \cite{hayward} and references therein). Following the leading work by Morris and Thorne \cite{mo} in which their main aspects were introduced, traversable Lorentzian wormholes have received considerable attention \cite{visser}. If traversable wormholes  exist, they could imply some unusual consequences, as for example the possibility of time travel \cite{mo-fro}. However, difficulties with their mechanical stability and the requirement of exotic matter (matter not fulfilling the energy conditions), which seems unavoidable within the framework of General Relativity, constitute the main 
objections against the actual existence of these objects. In particular,  these two central aspects have been thoroughly analyzed for wormholes of the thin-shell class \cite{povi,visser}, that is, geometries which are mathematically constructed by cutting and pasting two manifolds, so that a matter layer is located at the throat (see Ref. \cite{ernesto} and references therein).

About two decades ago, the geometrical aspects of topological defects like cosmic strings began to be the object of a detailed study, mainly because they could have played an important role in structure formation in the early universe; it was also pointed out that they could manifest by gravitational lensing effects (see Ref. \cite{vilenkin}). From a different point of view,  one-dimensional (open or closed) strings are the objects more seriously considered by present theoretical developments to be the fundamental building blocks of nature.  As a consequence, the interest in the gravitational effects of both fundamental and cosmic strings has been recently renewed  (for example, see Refs. \cite{strings}). It is thus natural that cylindrically symmetric wormhole geometries, as those associated to cosmic strings are, have been considered in the last years. 

Recent works in which  cylindrical wormholes have been studied include those  by Cl\'ement \cite{cle1,cle2}, Aros and Zamorano \cite{arza}, Kuhfittig \cite{ku}, Bronnikov and Lemos \cite{brle} and our papers \cite{ei1,ei2,mc}. The  thin-shell wormhole configurations associated to local and global cosmic strings  analyzed in Refs. \cite{ei1,ei2,mc} turned to be unstable under velocity perturbations conserving the cylindrical symmetry. Moreover, it was  pointed out \cite{ei2} that this feature seemed to be independent of the particular geometry considered, as long as the symmetry and the form of the equations of state of the static configurations were preserved. Mechanical instability would constitute an additional restriction to the possibility of finding these objects in the present day universe, besides the usual problem with exotic matter. However, regarding the issue of exotic matter, there is a subtle point in the case of non compact wormholes --as cylindrically symmetric ones are--. It was noted in Ref. 
\cite{brle} that for such configurations there are two admissible definitions of the throat, and that one of them --that the geodesics {\it restricted to a plane normal to the symmetry axis} open up-- can be compatible with a positive energy density.

In this work we extend our previous analysis of non-rotating cylindrical thin-shell wormholes. We first construct them by cut and paste of two equal regions of the most general static cylindrically symmetric geometry. Then we consider small velocity perturbations, under the hypothesis that the form of the equations of state of matter on the shell are preserved in the subsequent evolution. Our results demonstrate the conjecture about the stability of these objects introduced in Ref. \cite{ei2}. Second, we discuss the implications of relaxing the flare-out condition, as proposed in \cite{brle}. We find that within this new approach, for certain values of the  parameters characterizing the configurations, cylindrical thin-shell wormholes could be supported by matter of positive energy density. The analysis is exemplified with wormholes constructed starting from Einstein--Maxwell cylindrical geometries. Throughout the paper we adopt units such $G=c=1$.

\section{Wormhole construction and characterization}\label{cylind}

In coordinates $X^{\alpha}=(t,r,\varphi ,z)$, the general static  metric with cylindrical symmetry can be written in the form 
\begin{equation}
ds^2 = -A(r)dt^2 +B(r)dr^2 +C(r)d\varphi ^2+D(r)dz^2,
\label{e1}
\end{equation}
where $A$, $B$, $C$ and $D$ are positive functions of $r$. From this geometry, we take two copies $ \mathcal{M}^{\pm} = \{ x / r \geq a \}$ of the region with $r \geq a$ and join them at the hypersurface
$ \Sigma \equiv \Sigma^{\pm} = \{ x / r - a = 0 \}$ to make a manifold $\mathcal{M}=\mathcal{M}^{+} \cup \mathcal{M}^{-}$, which is geodesically complete. If the geometry opens up at the shell (flare-out condition), this construction creates a cylindrically symmetric thin-shell wormhole with two regions connected by a throat at $\Sigma $. For the analysis of this traversable wormhole, we use the standard Darmois-Israel formalism \cite{daris} (for a modern review see Ref. \cite{mus}).

The throat of the wormhole is a synchronous timelike hypersurface, in which we adopt coordinates $\xi ^i=(\tau , \varphi,z )$, with $\tau $ the proper time on the shell. For the subsequent analysis of the stability of the static configurations, we let the radius of the throat be a function of $\tau $, i.e. $a = a ( \tau )$. In this case, the shell is defined by $\Sigma : \mathcal{H} ( r, \tau ) = r - a ( \tau ) = 0$. The extrinsic curvature (second fundamental forms) associated with the two sides of the shell is then given by
\begin{equation}
  K_{ij}^{\pm} = - n_{\gamma}^{\pm} \left. \left( \frac{\partial^2
  X^{\gamma}}{\partial \xi^i \partial \xi^j} +\Gamma_{\alpha\beta}^{\gamma}
  \frac{\partial X^{\alpha}}{\partial \xi^i} \frac{\partial
  X^{\beta}}{\partial \xi^j} \right) \right|_{\Sigma}, \label{e2}
\end{equation}
where $n_{\gamma}^{\pm}$ are the unit normals ($n^{\gamma} n_{\gamma} = 1$) to
$\Sigma$ in $\mathcal{M}$:
\begin{equation}
  n_{\gamma}^{\pm} = \pm \left| g^{\alpha \beta} \frac{\partial
  \mathcal{H}}{\partial X^{\alpha}} \frac{\partial \mathcal{H}}{\partial
  X^{\beta}} \right|^{- 1 / 2} \frac{\partial \mathcal{H}}{\partial
  X^{\gamma}} . \label{e3}
\end{equation}
It is convenient to use the orthonormal basis $\{ e_{\hat{\tau}} = \sqrt{1/A(r)}e_{t}$, $e_{\hat{\varphi}} 
= \sqrt{1/C(r)}e_{\varphi}$, $e_{\hat{z}} =\sqrt{1/D(r)}e_{z}\}$, 
for which $g_{_{\hat{\imath} \hat{\jmath}}} = \eta_{_{\hat{\imath}
\hat{\jmath}}}=diag(-1,1,1)$, so we obtain
\begin{equation}
K_{\hat{\tau} \hat{\tau}}^{\pm} = \mp \frac{2A(a)B(a) \ddot{a}+A'(a)
+[A(a)B'(a)+A'(a)B(a)]\dot{a}^2}{2A(a)\sqrt{B(a)} \sqrt{1+B(a)\dot{a}^2}},
\label{e4}
\end{equation}
\begin{equation}
K_{\hat{\varphi} \hat{\varphi}}^{\pm} = \pm \frac{C'(a)\sqrt{1+B(a)
\dot{a}^2}}{2C(a)\sqrt{B(a)}}, 
\label{e5}
\end{equation}
and
\begin{equation}
K_{\hat{z} \hat{z}}^{\pm} = \pm \frac{D'(a)\sqrt{1+B(a)\dot{a}^2}}
{2D(a) \sqrt{B(a)}}, 
\label{e6}
\end{equation}
where the dot and the prime represent $d / d \tau$ and $d/dr$, respectively. The Einstein equations on the shell, also called the Lanczos equations, can be written in the form:
\begin{equation}
-[K_{\hat{\imath} \hat{\jmath}}]+[K]g_{\hat{\imath} \hat{\jmath}}=
8\pi S_{\hat{\imath} \hat{\jmath}},
\label{e7}
\end{equation}
where $[K_{_{\hat{\imath} \hat{\jmath}}}]\equiv K_{_{\hat{\imath}
\hat{\jmath}}}^+ - K_{_{\hat{\imath} \hat{\jmath}}}^-$, 
$[K]=g^{\hat{\imath} \hat{\jmath}}[K_{\hat{\imath} \hat{\jmath}}]$ is the 
trace of $[K_{\hat{\imath} \hat{\jmath}}]$ and
$S_{_{\hat{\imath} \hat{\jmath}}} = \text{\textrm{diag}} ( \sigma, 
p_{\varphi }, p_{z} )$ is the surface stress-energy tensor, with $\sigma$ the surface energy density and $p_{\varphi , z}$ the surface pressures. By replacing Eqs. (\ref{e4}), (\ref{e5}) and (\ref{e6}) in Eq. (\ref{e7}) we have
\begin{equation}
  \sigma = - \frac{\sqrt{1 + B(a) \dot{a}^2}}{8 \pi \sqrt{B(a)}} \left[
  \frac{C'(a)}{C(a)} + \frac{D'(a)}{D(a)} \right],
\label{e8}
\end{equation}
\begin{equation}
p_{\varphi} =  \frac{1}{8 \pi \sqrt{B(a)} \sqrt{1 + B(a)\dot{a}^2}} 
\left\{ 2 B(a) \ddot{a} + B(a) \left[ \frac{D'(a)}{D(a)} + \frac{A'(a)}{A(a)} +
\frac{ B'(a)}{B(a)} \right] \dot{a}^2 + \frac{D'(a)}{D(a)} + \frac{A'(a)}{A(a)} \right\},
\label{e9}
\end{equation}
\begin{equation}
p_{z} =  \frac{1}{8 \pi \sqrt{B(a)} \sqrt{1 + B(a) \dot{a}^2}}
\left\{ 2 B(a) \ddot{a} + B(a) \left[ \frac{C'(a)}{C(a)} + \frac{A'(a)}{A(a)} + \frac{B'(a)} {B(a)} \right] \dot{a}^2 + \frac{C'(a)}{C(a)} + \frac{A'(a)}{A(a)} \right\} .
\label{e10}
\end{equation}
It is easy to see that $p_{\varphi }$, $p_{z}$ and $\sigma$ satisfy the equation
\begin{equation}
p_z-p_{\varphi }=\frac{C(a)D'(a)-C'(a)D(a)}
{C(a)D'(a)+C'(a)D(a)}\sigma .
\label{e11}
\end{equation}

The static equations are obtained by putting $\dot{a}=0$ and $\ddot{a}=0$ in Eqs. 
(\ref{e8}), (\ref{e9}) and (\ref{e10}):
\begin{equation}
\sigma = - \frac{1}{8 \pi \sqrt{B(a)}} \left[
\frac{C'(a)}{C(a)} + \frac{D'(a)}{D(a)} \right],
\label{e12}
\end{equation}
\begin{equation}
p_{\varphi} =  \frac{1}{8 \pi \sqrt{B(a)} } 
\left[\frac{D'(a)}{D(a)}+ \frac{A'(a)}{A(a)} \right],
\label{e13}
\end{equation}
\begin{equation}
p_{z} =  \frac{1}{8 \pi \sqrt{B(a)}} 
\left[\frac{C'(a)}{C(a)} + \frac{A'(a)}{A(a)} \right] .
\label{e14}
\end{equation}
By using Eq. (\ref{e12}), the Eqs. (\ref{e13}) and (\ref{e14}) can be recast in the form
\begin{equation}
p_{\varphi}=-\frac{C(a)[A(a)D'(a)+A'(a)D(a)]}{A(a)[C(a)D'(a)+C'(a)D(a)]} \sigma,
\label{e15}
\end{equation}
\begin{equation}
p_{z}=-\frac{D(a)[A(a)C'(a)+A'(a)C(a)]}{A(a)[C(a)D'(a)+C'(a)D(a)]} \sigma .
\label{e16}
\end{equation}
Then, the functions $A$, $C$ and $D$ determine the equations of state 
$p_{\varphi}(\sigma)$ and $p_{z}(\sigma)$ of the exotic matter on the shell.

If $\lp CD \rp'(a)=C'(a)D(a)+C(a)D'(a)>0$, from Eq. (\ref{e12}) it follows that the surface energy density is negative, indicating the presence of \textit{exotic} matter at the throat.  The usual definition of the wormhole throat --suitable for compact configurations-- states that it is  a {\it minimal area surface}. In the cylindrical case one can define the area function ${\cal A}(r)=\sqrt{C(r)D(r)}$, which should increase at both sides of the throat; then ${\cal A}'(a)>0$, so that $\lp CD \rp'(a)>0$. We shall call it the {\it areal} flare-out condition, for which negative energy density at the throat  is clearly  unavoidable. However,  Bronnikov and Lemos \cite{brle} recently pointed out that, for infinite cylindrical configurations,  a less restrictive  definition of the wormhole throat seems to be the most natural. This new definition  requires that  the circular radius function  ${\cal R}(r)=\sqrt{C(r)}$ has a minimum at the throat\footnote{Wormholes have a non trivial topology, which constitutes a 
global property of spacetime. For  static cylindrically symmetric geometries, the global properties are determined by the behavior of the circular radius function \cite{brle}.}, so this function should increase at both sides of it (the geodesics within a plane normal to the symmetry axis open up at the throat); thus we shall call this one  the {\it radial} flare-out condition. This definition implies $C'(a)>0$, leaving free the sign of $(CD)'(a)$; this feature is of central relevance for the physical viability of the wormhole, because allows for the possibility of a {\it positive} energy density\footnote{In our paper \cite{ei1} the less restrictive flare-out condition was adopted, while in  \cite{ei2} and \cite{mc} we worked with the other definition.}.

\section{Stability analysis}

In what follows, we assume that the equations of state for the dynamic case have the same form as in the static one, i.e. that they do not depend on the derivatives of $a(\tau)$, so $p_{\varphi}(\sigma)$ and $p_{z}(\sigma)$ are given by Eqs. (\ref{e15}) and (\ref{e16}). This assumption is justified by the fact that we are interested in small velocity perturbations starting from a static  solution, so that the evolution of the shell matter can be considered as a succession of static states \cite{ei2,mc}. Then, replacing Eqs. (\ref{e8}) and (\ref{e9}) in Eq. (\ref{e15}) (or Eqs. (\ref{e8}) and (\ref{e10}) in Eq. (\ref{e16})), a simple second order differential equation for $a(\tau )$ is obtained:
\begin{equation} 
2B(a)\ddot{a}+B'(a)\dot{a}^{2}=0.
\label{e17}
\end{equation}
It is easy to verify that 
\begin{equation} 
\dot{a}(\tau )=\dot{a}(\tau _{0})\sqrt{\frac{B(a(\tau _{0}))}{B(a(\tau))}},
\label{e18}
\end{equation}
satisfies Eq. (\ref{e17}), where $\tau _{0}$ is an arbitrary fixed time.
Then, Eq. (\ref{e18}) can be rewritten in the form
\begin{equation} 
\sqrt{B(a)}da=\dot{a}(\tau _{0})\sqrt{B(a(\tau _{0}))}d\tau ,
\label{e19}
\end{equation}
which by integrating both sides gives
\begin{equation} 
\int^{a(\tau )}_{a(\tau _{0})}\sqrt{B(a)}da=\dot{a}(\tau _{0})
\sqrt{B(a(\tau _{0}))}(\tau - \tau _{0}).
\label{e20}
\end{equation}
So we have that the time evolution of the radius of the throat $a(\tau )$ is formally obtained by calculating the integral and inverting Eq. (\ref{e20}). From Eq. (\ref{e18}) one concludes that the sign of the velocity of the shell after being perturbed is completely determined by the sign of the initial velocity. Depending on the fact that the metric function $B$ is an increasing or a decreasing function of the throat radius, for a positive initial velocity the absolute value of the velocity will respectively decrease or increase, while for a negative initial velocity the opposite result is obtained. No oscillatory behavior can take place: the shell can only undergo a monotonous evolution. Because we have started from the most general static cylindrically symmetric geometry, we then find that the conjecture introduced in Ref. \cite{ei2} is true: under the assumption that the symmetry is preserved and that the equations of state of matter on the shell corresponding to  static configurations are kept valid 
after a perturbation, cylindrical thin-shell wormholes are unstable.

\section{Examples: Einstein--Maxwell spacetimes}

There exist three kinds of static cylindrically symmetric geometries associated to Maxwell electromagnetism coupled to Einstein gravity \cite{ks}: with the definition $G(r)=k_1r^m+k_2r^{-m}$, we have the metric associated to an axial current (angular magnetic field)
\be
ds^2=r^{2m^2}G^2(r)\lp-dt^2+dr^2\rp+r^2G^2(r)d\varphi^2+G^{-2}(r)dz^2,
\ee
the metric corresponding to an angular current (magnetic field in the direction of the symmetry axis)
\be
ds^2=r^{2m^2}G^2(r)\lp-dt^2+dr^2\rp+G^{-2}(r)d\varphi^2+r^2G^2(r)dz^2,
\ee
and the metric associated to an axial charge (radial electric field):
\be
ds^2=-G^{-2}(r)dt^2+r^{2m^2}G^2(r)\lp dr^2+dz^2\rp+r^2G^2(r)d\varphi^2.\label{E}
\ee
The constants $k_1$, $k_2$ and $m$ are real numbers that must fulfill $k_1k_2 > 0$ in the magnetic cases and $k_1k_2 < 0$ in the electric case, so the metric and the electromagnetic potential are real \cite{ks}. The coordinates adopted above have been suitably adimensionalized in order to avoid constants which do not play an important physical role. 

\begin{figure}[t!]
\begin{center}
\vspace{0cm}
\includegraphics[width=15cm]{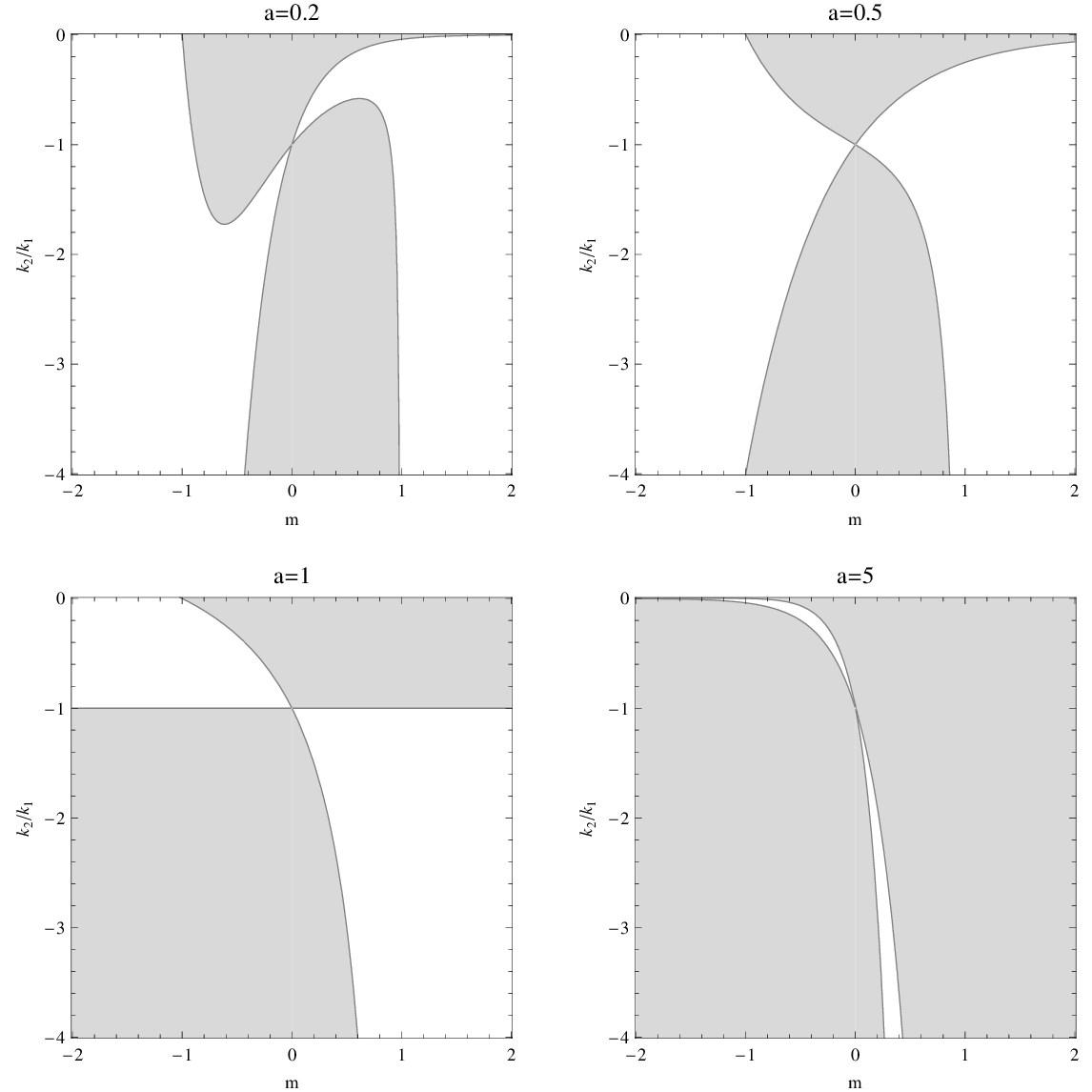}
\vspace{0cm}
\end{center} 
\caption{Wormholes of throat radius $a$ with a radial electric field: the gray zones correspond to the values of the parameters $m$, $k_1$ and $k_2$ for which the radial flare-out condition is satisfied.} 
\label{fig1}
\end{figure}
\begin{figure}[t!]
\begin{center}
\vspace{0cm}
\includegraphics[width=15cm]{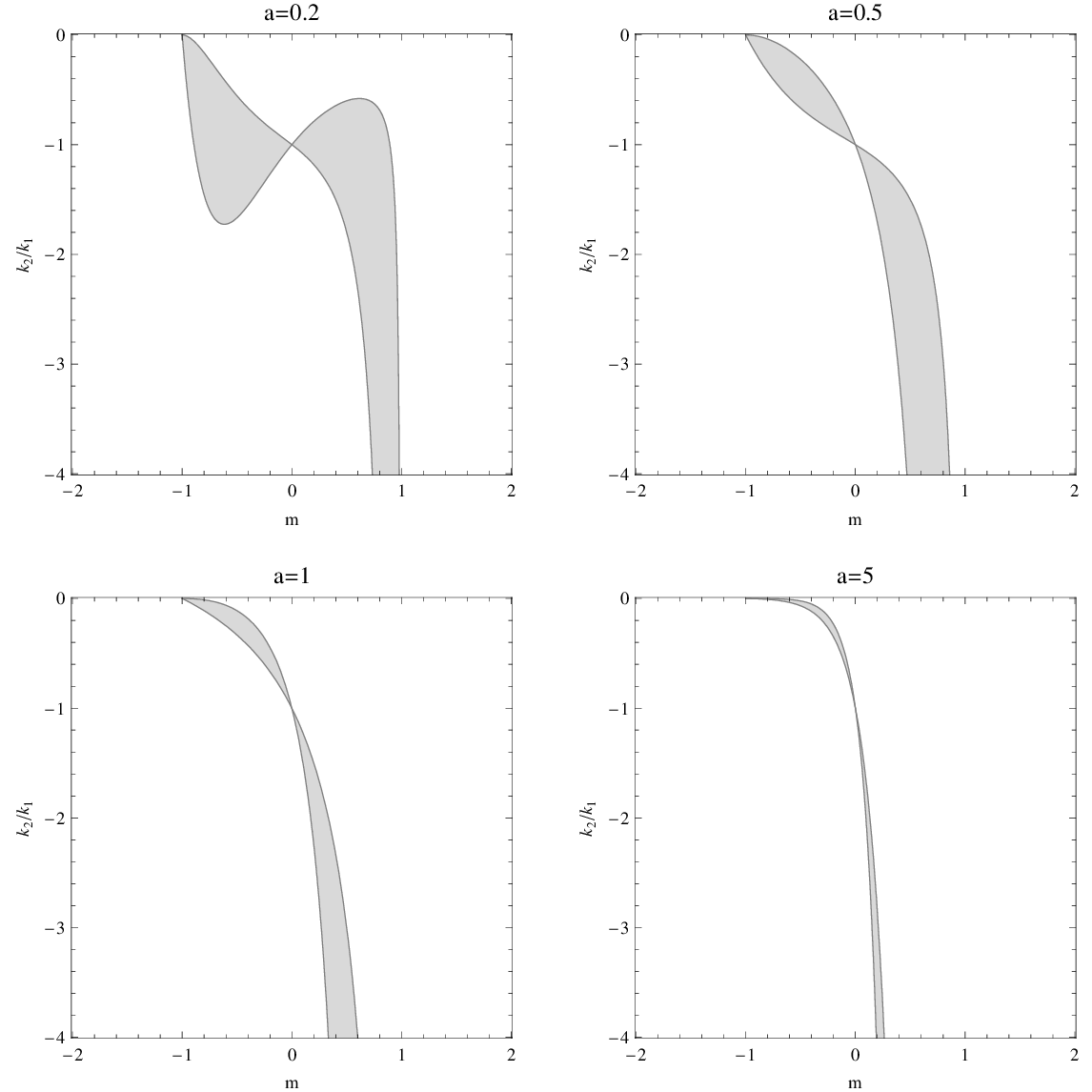}
\vspace{0cm}
\end{center} 
\caption{Wormholes of throat radius $a$ with a radial electric field: the gray zones correspond to the values of the parameters $m$, $k_1$ and $k_2$  for which the radial flare-out condition is satisfied and the energy density at the throat is positive.} 
\label{fig2}
\end{figure}

In terms of the functions of the general static metric (\ref{e1}), in the first magnetic case we have $A(r)=B(r)=G^2(r)r^{2m^2}$, $C(r)=r^2G^2(r)$, $D(r)=G^{-2}(r)$, in the second magnetic case  $A(r)=B(r)=G^2(r)r^{2m^2}$, $C(r)=G^{-2}(r)$, $D(r)=r^2G^2(r)$, and in the electric case $A(r)=G^{-2}(r)$, $B(r)=D(r)=r^{2m^2}G^2(r)$, $C(r)=r^2G^2(r)$.  Thus  in both magnetic cases we obtain the same behaviour for the product $CD$, whose radial derivative, after the mathematical construction of a wormhole of throat radius $a$,  determines the sign of the surface energy density $\sigma$: $C(r)D(r)=r^2$, so that $\lp CD\rp'(a)=2a >0$ and $\sigma$ can only be negative. In the case with a radial electric field, instead,  we have $C(r)D(r)=r^{2m^2+2}G^4(r)$, which not necessarily is an increasing function of $r$. Thus we can have $\lp CD \rp' (a)<0$ and, if this is compatible with $ C'(a)>0$, working with the radial flare-out condition a cylindrical thin-shell wormhole would be possible with $\sigma >0$. For the metric (\ref{E}) associated to the electric field we have
\begin{eqnarray}
C'(a)& = & 2aG(a)\lb G(a)+aG'(a)\rb \nonumber\\
& = & 2a\lp k_1a^m+k_2a^{-m}\rp \lb (1+m)k_1 a^m+(1-m)k_2a^{-m}\rb,\label{geo}
\end{eqnarray}
\begin{eqnarray}
\lp CD\rp '(a)& = & 2a^{2m^2+1}G^3(a)\lb(m^2+1)G(a)+2aG'(a)\rb\nonumber\\
& = & 2a^{2m^2+1}\lp k_1a^m+k_2a^{-m}\rp^3\lb (1+m)^2 k_1a^m+(1-m)^2k_2a^{-m}\rb.\label{ener}
\end{eqnarray}
If values of the parameters $k_1$, $k_2$ and $m$ exist such that  the expression  (\ref{geo}) is positive definite and the expression (\ref{ener}) is negative, then there exist  cylindrical thin-shell wormholes of throat radius $a$  which are supported by matter of positive energy density. Let us see for which values of the parameters we obtain $\sigma > 0$. Eq. (\ref{ener}) can be written in the form
 \be
\lp CD\rp '(a) =  2a^{2m^2+1}G^2(a)\left\{2G(a)\lb G(a)+ aG'(a)\rb+(m^2-1)G^2(a)\right\}.
\ee
The first term within the curly brackets is always positive if $C'(a)>0$; then a necessary (but not sufficient) condition for $\lp CD\rp '(a)<0$ is that $|m|<1$. The values of the parameters $m$, $k_1$ and $k_2$ for which the radial flare-out condition is satisfied are plotted in Fig. \ref{fig1}, and the values for which the energy density is positive are shown in Fig. \ref{fig2}, in both cases for different values of the wormhole throat radius $a$.  

Now let us analyze the  energy conditions: non exotic matter satisfies the weak energy condition (WEC) that states $\rho \geq 0$, $\rho+P_\mu\geq 0$ ($\mu=r,\varphi,z$), or at least the null energy condition (NEC) $\rho+P_\mu\geq 0$. In our  thin-shell construction the energy conditions are satisfied outside the shell. On the throat, instead of the energy density $\rho$  we have the surface energy density $\sigma$,  and instead of the pressures $P_\mu$ we have the surface pressures $p_{\varphi}$, $p_z$ and $p_r$. It is straightforward to see that 
\be
p_z=\frac{1}{4\pi a^{m^2+1}|k_1a^m+k_2a^{-m}|}\qquad \mathrm{and} \qquad
p_{\varphi}  =  \frac{m^2}{4\pi a^{m^2+1}|k_1a^m+k_2a^{-m}|}.
\label{spf}
\ee
So we find that $p_{z}> 0$ and $p_{\varphi}\geq 0$, for all the values of the parameters. On the other hand, $\sigma+p_r=\sigma$ because the radial  pressure is non singular on the shell. Then it is possible to satisfy the energy conditions when the energy density is positive. If the parameters are such that $\sigma >0$, then an observer at rest at the throat, or an observer moving from one side of the throat to the other side would find only positive energy matter.

\section{Summary}\label{discu}

We have constructed cylindrical thin shell wormholes of a general class, symmetric with respect to the throat. Under the assumption that the equations of state on the shell which defines the throat have the same form as in the static case when it is perturbed preserving the symmetry, we have shown that these wormholes are always mechanically unstable. We have therefore demonstrated our conjecture presented in \cite{ei2}: this is a general property and does not depend on the particular form of the metric adopted for the construction. Two possible definitions of the flare-out condition were considered. The areal flare-out condition is more restrictive because it always leads to a negative energy density on the throat. Instead, in this work we have shown that the radial one is less restrictive, allowing for positive energy density, depending on the particular form of the original metric from which the construction was performed. Examples of wormholes with electromagnetic fields were presented. In the two cases with a magnetic field, the energy density is always negative for both definitions of the flare-out condition. For a radial electric field, with the adoption of the radial flare-out condition we have found values of the parameters such that the energy density at the throat is positive. In this case, we have shown that the weak energy condition is also satisfied.

\section*{Acknowledgments}

This work has been supported by Universidad de Buenos Aires and CONICET. In the previous versions of the article, there was an algebraic error in Eq. (\ref{e4}), which propagates to a few equations without affecting the main results. We thank Cecilia Bejarano for pointing this out.

\end{document}